## SE FOR AI

# Love, Joy, Anger, Sadness, Fear, and Surprise

## SE Needs Special Kinds of AI: A Case Study on Text Mining and SE

Nicole Novielli, Fabio Calefato, and Filippo Lanubile

**From the Editor**
Artificial-intelligence (AI) tools are often applied to software engineering (SE) tasks using their "off-the-shelf" configurations. But is that wise? Perhaps not. In this column, researchers from the University of Bari show how AI gets much better when it is tuned to the particulars of SE.—*Tim Menzies*

**DO YOU LIKE** your code? What kind of code makes developers happiest? What makes them angriest? Is it possible to monitor the mood of a large team of coders to determine when and where a codebase needs additional help?

Answering these questions is the task of text mining or, more specifically, sentiment analysis. Sentiment analysis is the automatic processing of comments to map and capture the polarity of emotions and opinions. Much recent work has explored sentiment analysis in software engineering (SE) using ready-made artificial-intelligence (AI) tools to mine emotions and opinions in collaborative development platforms and app stores. The use of sentiment analysis in SE is greatly facilitated by the maturity of its methodologies and techniques, which has resulted in a plethora of publicly available tools.

To assess the merits of sentiment analysis, the case studies of this article look at six basic emotions, namely: love, joy, anger, sadness, fear, and surprise, as conveyed in questions, answers, and comments about code within two data sources. First, we look at the Stack Overflow programmer discussion site. Second, we consider the Jira site. For all of these data, the goal is to explore more than 10,000 questions, answers, and comments, annotated by emotion polarity.

Based on the data, we write to warn against using publicly available AI/machine-learning tools off the shelf (i.e., without modification).[1] Such tools have been trained in domains that are far from software development. Jongeling et al.[2] have shown how general-purpose sentiment-analysis tools produce unreliable results in SE. We have shown[3] that uncustomized sentiment-analysis tools perform worse than tools (described in "SE-Specific Sentiment Analysis Tools") tuned to the specifics of SE data.

### SE Needs Special Kinds of AI

#### Does SE-Specific Tuning Enhance Sentiment Analysis?
To answer this question, we evaluated the performances of the Senti4SD,

SentiCR, and SentiStrengthSE classifiers on two publicly available data sets specifically built to investigate the role of emotions in software development, the Stack Overflow data set, which was originally used for training Senti4SD, and the Jira data set by Ortu and colleagues.[4] Both data sets have been annotated following a model-driven approach: the human raters were trained to label the text to capture the polarity of six basic emotions: love, joy, anger, sadness, fear, and surprise.

The Stack Overflow data set includes 4,423 questions, answers, and comments annotated by emotion polarity. The Jira data set includes approximately 2,000 issue comments and 4,000 sentences from popular open source software projects, such as Apache, Spring, JBoss, and CodeHaus. We split each data set into train (70%) and test (30%) sets. We used the train set to retrain the supervised classifiers Senti4SD and SentiCR, and we then assessed the performance on the test set.

To enable a fair comparison, we used the same test partition to assess SentiStrengthSE, as it leverages a lexicon-based approach and cannot be retrained. As a baseline, we refer to the performance of SentiStrength, a general-purpose tool widely used in social computing research.

We measured performance in terms of precision, recall, and F-measure for all three polarity classes (see Table 1). We also report the average precision, recall, and F1-measure to enable a quick comparison of the overall performance of each classifier. In addition, we report the weighted Cohen kappa ($\kappa$) to measure both the agreement with gold labels and the agreement among the three tools (see Table 2). We distinguish between mild disagreement (between negative/positive and neutral annotations) and strong disagreement (between positive and negative judgments). As such, we assigned a weight of two and one to strong and mild disagreement, respectively, to compute the weighted $\kappa$. The agreement is considered substantial if $0.61 \leq \kappa \leq 0.80$ and almost perfect if $0.81 \leq \kappa \leq 1$. For the sake of completeness, we also provide the percentage of observed agreement.

The results show that the SE-specific tools outperform the general-purpose baseline (SentiStrength), thus suggesting that the customization of sentiment-analysis tools to the software development domain successfully improves the classification accuracy (see Tables 1 and 2). Specifically, the high recall/low precision for negative class and the high precision/low recall for neutral class observed for SentiStrength on the Stack Overflow data set indicate a bias of general-purpose tools toward the recognition of negative sentiments. This is mainly due to SentiStrength misclassifying sentences using neutral technical lexicon as negative: for example, "I want to kill this process" is erroneously labeled as negative.

### Table 1. The performance of sentiment-analysis tools for model-driven annotations.

| Data set | Class | SentiStrength (baseline) | | | Senti4SD | | | SentiStrengthSE | | | SentiCR | | |
|---|---|---|---|---|---|---|---|---|---|---|---|---|---|
| | | P | R | F1 | P | R | F1 | P | R | F1 | P | R | F1 |
| Stack overflow | Positive | 0.89 | **0.92** | 0.9 | **0.92** | **0.92** | **0.92** | 0.89 | 0.83 | 0.86 | 0.88 | 0.9 | 0.89 |
| | Negative | 0.67 | **0.96** | 0.79 | **0.8** | 0.89 | **0.84** | 0.75 | 0.79 | 0.77 | 0.79 | 0.73 | 0.76 |
| | Neutral | **0.95** | 0.64 | 0.76 | 0.87 | 0.8 | **0.83** | 0.75 | 0.77 | 0.76 | 0.79 | **0.82** | 0.8 |
| | Overall | 0.84 | 0.84 | 0.82 | **0.86** | **0.87** | **0.86** | 0.8 | 0.8 | 0.8 | 0.82 | 0.82 | 0.82 |
| Jira | Positive | 0.5 | **0.91** | 0.65 | **0.76** | 0.79 | 0.78 | 0.69 | 0.94 | 0.8 | **0.76** | 0.89 | **0.82** |
| | Negative | 0.41 | 0.64 | 0.5 | 0.72 | 0.57 | 0.64 | 0.67 | 0.71 | 0.69 | **0.81** | 0.61 | **0.7** |
| | Neutral | 0.89 | 0.59 | 0.71 | 0.86 | 0.89 | 0.88 | **0.92** | 0.82 | 0.87 | 0.89 | **0.89** | **0.89** |
| | Overall | 0.6 | 0.71 | 0.62 | 0.78 | 0.75 | 0.77 | 0.76 | **0.82** | 0.79 | **0.82** | 0.8 | **0.8** |

*The values highlighting the best performances are given in bold.*

Table 2. The agreement of SE-specific tools with manual labeling and with each other for model-driven annotations.

| Manual labeling | | | | | Model-driven annotation | | | | |
|---|---|---|---|---|---|---|---|---|---|
| | Agreement metrics | | | | | Agreement metrics | | | |
| | | Agreement (%) | Disagreement (%) | | Classifiers | k | Agreement (%) | Disagreement (%) | |
| Classifier | k | | Severe | Mild | | | | Severe | Mild |
| Stack Overflow | | | | | | | | | |
| Senti4SD | 0.83 | 86 | 1 | 12 | Senti4SD versus SentiCR | 0.77 | 83 | 3 | 14 |
| SentiStrengthSE | 0.74 | 80 | 2 | 18 | Senti4SD versus SentiStrengthSE | 0.79 | 84 | 2 | 15 |
| SentiCR | 0.76 | 82 | 3 | 15 | SentiCR versus SentiStrengthSE | 0.73 | 80 | 3 | 17 |
| SentiStrength | 0.77 | 82 | 3 | 15 | | | | | |
| Jira | | | | | | | | | |
| Senti4SD | 0.67 | 83 | 0 | 17 | Senti4SD versus SentiCR | 0.76 | 88 | 0 | 12 |
| SentiStrengthSE | 0.7 | 83 | 0 | 17 | Senti4SD versus SentiStrengthSE | 0.7 | 83 | <1 | 16 |
| SentiCR | 0.73 | 86 | 0 | 14 | SentiCR versus SentiStrengthSE | 0.81 | 89 | <1 | 10 |
| SentiStrength | 0.48 | 66 | <2 | 33 | | | | | |

*SentiStrength is included as a baseline. The percentages may not add up to 100% due to rounding.*

This bias is corrected by the semantic features in Senti4SD. In the presence of unbalanced training data, the use of resampling techniques is also beneficial. In the case of the Jira data set (68% of neutral cases), we observe the best performance using SentiCR, probably due to the synthetic minority oversampling technique optimization it performs on the training set.

Another issue with general-purpose tools is that they do not agree with each other, thus making sentiment analysis tool dependent when applied to SE. Conversely, the $\kappa$ agreement for SE-specific tools ranges from substantial to perfect for all couples of tools (see Table 2). Mild disagreement is the main cause of the difference in performance. Strong disagreement is never observed on the Jira data set and is equal to only 3% in the worst case, indicating the general reliability of SE-specific tools.

### Do We Need a Theoretical Model of Affect?

When using ready-made sentiment-analysis tools, we run the risk that a mismatch exists between the tool builders and our goals. Indeed, the implementation choices made by tool designers are driven by the intended use of the classification output. Let us consider, for example, the sentence "This method is slow, and I would not recommend using it." This text conveys a negative opinion, which might be relevant if our goal is to build a recommender system. Conversely, if our focus is on emotions, such as for community moderation purposes or for early intervention to prevent burnout, the very same comment should be labeled as neutral because it does not convey any emotional content.

In fine-tuning sentiment-analysis tools, we need to keep in mind the role played by the theoretical model of affect in preparing the gold standard for the training, in the case of supervised classifiers, or in fine-tuning the lexical resources, in case of rule-based approaches. To investigate the impact of referring to a clear conceptualization of affect on the final performance, we replicate the benchmarking study using two additional data sets for which

Table 3. The performance of sentiment-analysis tools on ad-hoc annotation.

| Data set | SentiStrength (baseline) | | | Senti4SD | | | SentiStrengthSE | | | SentiCR | | |
|---|---|---|---|---|---|---|---|---|---|---|---|---|
| | P | R | F1 | P | R | F1 | P | R | F1 | P | R | F1 |
| Code Review | 0.58 | 0.57 | 0.57 | 0.75 | 0.7 | 0.69 | 0.66 | 0.59 | 0.6 | **0.76** | **0.76** | **0.76** |
| Java Libraries | 0.42 | 0.42 | 0.41 | **0.67** | 0.51 | 0.56 | 0.46 | 0.4 | 0.4 | 0.62 | **0.58** | **0.59** |

*Values highlighting the best performances are given in bold.*

ad hoc annotation was performed: the corpus of code reviews used for training SentiCR (hereafter called *Code Review*) and a Java Libraries data set.[5]

These data sets were labeled by simply asking the raters to annotate the positive, negative, or neutral semantic orientation of a text based on their subjective perception and without any further guidance or training. The Code Review corpus includes 2,000 review comments from 20 popular open source software projects using Gerrit. The Java Libraries data set was collected in the scope of a broader study aiming at developing a recommender for software libraries that leverages sentiment analysis for mining crowd-sourced opinions. It includes 1,500 sentences randomly extracted from Stack Overflow.

We repeated the same train test setting implemented for the Stack Overflow and Jira data sets (model driven). In Table 3, we report the overall classification performance in terms of average precision, recall, and F1-measure. Compared to the model-driven annotation, we observe a drop in performance for the ad hoc annotation data sets. The lower performance on the ad hoc annotation data sets is also reflected in values of $\kappa$ ranging from slight to moderate agreement, with the highest values of $\kappa = 0.50$ observed for SentiCR on the Code Review comments. After performing an error analysis, we discovered that the main cause for misclassification for ad hoc annotation data sets is due to "polar facts," which are inherently desirable or not. Therefore, they usually invoke, for most people, a positive or negative feeling.

This is the case, for example, of a problem report for which the reported event is inherently negative but the emotional tone of the sentence is neutral, as in, "I tried the following and it returns nothing." Such cases are consistently annotated as neutral in the model-driven data sets; conversely, they are often labeled as either positive or negative in the ad hoc data sets, based on the subjective perception of the human rater, thus introducing noise into the training. This is the main cause for misclassification for ad-hoc-annotation data sets, with 61% and 89% of misclassified texts for Code Review and Java libraries, respectively.

Conversely, we observe this error only in 7% and 12% of misclassified texts for the Stack Overflow and Jira data sets, respectively. This evidence provides an indication of the importance of grounding sentiment analysis in a theoretical model of affect that clearly represents the phenomenon we would like to identify in text.

SE-specific sentiment-analysis tools represent a considerable step forward, compared to using general-purpose solutions for analyzing content from the software-development domain. Still, these are not "general enough" to be used *tout court,* as the intended use and the data set lexicon may vary. To make them usable for SE, we offer four easy-to-follow guidelines to make sure that researchers and practitioners (re)use existing sentiment-analysis solutions correctly and obtain reliable results.

- *Use SE-specific tools*: Reliable sentiment analysis in SE is possible provided that SE-specific tools are used. Customization of sentiment-analysis tools produces better classification accuracy versus general-purpose tools trained on corpora from different domains, such as news or social media. This is particularly useful for solving the negative bias of general-purpose tools, which tend to classify neutral sentences containing technical lexicon as negative.
- *Use training data from your target platform*: Supervised tools, namely, Senti4SD and SentiCR, outperform SentiStrengthSE, which implements a lexicon-based approach. However, the customization might

### SOFTWARE-ENGINEERING-SPECIFIC SENTIMENT-ANALYSIS TOOLS

**Senti4SD**

Senti4SD is a polarity classifier specifically trained to support sentiment analysis in developers' communication channels. It leverages a suite of features based on bag of words, sentiment lexicons, and semantic features based on word embeddings trained on more than 20 million technical texts from Stack Overflow. It is publicly available for research purposes and is distributed with a classification model trained and validated on a gold standard of about 4,000 questions, answers, and comments extracted from Stack Overflow. The Senti4SD toolkit provides a training method for customizing the classifiers using an annotated gold standard as input. The Stack Overflow gold standard is also distributed. Website: https://collab-uniba.github.io/EMTk/

**SentiCR**

SentiCR is a supervised sentiment-analysis toolkit, specifically trained and evaluated for code-review comments. SentiCR implements preprocessing to normalize text, handle negations and emoticons, remove stop words (i.e., articles, prepositions, conjunctions, and pronouns), derive word stems, and remove code snippets. Furthermore, it performs a synthetic minority oversampling technique to address the problem of class imbalance in the training data. SentiCR has been evaluated using eight supervised algorithms in a 10-fold cross-validation setting. The currently distributed version includes the original gold-standard data set of code-review comments and allows retraining. Website: https://github.com/senticr/SentiCR

**SentiStrengthSE**

SentiStrengthSE implements a set of heuristics leveraging the sentiment lexicons, that is, large collections of words annotated with their prior polarity (i.e., the positive or negative semantic orientation of the word). The overall sentiment of a text is computed based on the prior polarity of the words composing it under the assumption that words with negative prior polarity convey negative sentiment, and vice versa. Specifically, SentiStrengthSE is built on application programming interface of SentiStrength, a general-purpose tool widely used in social computing research. It implements a rule-based approach that assigns a score for both positive and negative sentiment to the input text. The scores are computed based on a manually adjusted version of the original SentiStrength lexicon, that is, lists of positive and negative words in the lexicon whose sentiment scores were manually adjusted to reflect the semantics and neutral polarity of technical words, such as *support* or *default*. SentiStrengthSE has been optimized using a gold standard of manually annotated sentences from Jira. Website: https://laser.cs.uno.edu/Projects/Projects.html

---

produce a different performance on different data sources due to platform-specific jargon and communication style. For example, the use of negative lexicon in Stack Overflow might be simply due to the intended use of the platform, which is reporting problems. Similarly, approval messages, such as "Looks good to me," might be expected on code-review platforms used to approve a code change. As such, we recommend retraining supervised tools using a gold standard from the same domain and data source being targeted. Although manually labeling data sets is costly, Senti4SD produces optimal performance even with a minimal set of training documents.[6]

- *If you cannot retrain, perform a sanity check*: When a manually annotated gold standard is not available for retraining, unsupervised tools are the only option. In such cases, we recommend testing the selected tool on a representative sample of your target data and manually investigate the output to assess if it is in line with your intended use. In our benchmark study, we observed that the performance of SentiStrengthSE is comparable to that of supervised tools (see Table 1).

- *Define your goal first, then select the tool*: One of the most dangerous hidden assumptions when reusing AI components is that, by reusing tools and data sets, you are also implicitly inheriting the original goals and conceptualization of sentiment of its creators. In fact, based on the specific goals addressed, one might be interested in detecting specific emotions (e.g., frustration, anger, sadness, joy, satisfaction), opinions (i.e., positive or negative evaluations), or interpersonal stances (i.e., friendly versus hostile attitude).

The comparison of performances between model-driven and ad hoc annotated data sets shows that retraining on SE data sets still might not be enough to guarantee a satisfactory accuracy for all polarity classes when ad hoc annotation is adopted for building the gold standard. This provides further evidence that good training data are as important as good tools and that the absence of clear guidelines for the annotation of sentiment leads to noisy gold standards. Thus, we underline the need to distinguish between the task of identifying the affective content conveyed by a text (i.e., sentiment analysis) and the task of identifying the objective report of (un)pleasant facts in the developers' comments.

**ABOUT THE AUTHORS**

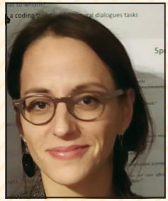

**NICOLE NOVIELLI** is an assistant professor with the University of Bari, Italy. Contact her at nicole.novielli@uniba.it.

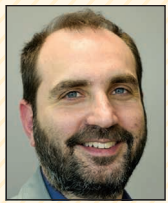

**FABIO CALEFATO** is an assistant professor with the University of Bari, Italy. Contact him at fabio.calefato@uniba.it.

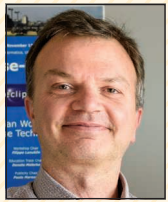

**FILIPPO LANUBILE** is a full professor with the University of Bari, Italy. Contact him at filippo.lanubile@uniba.it.